# On complexity of special maximum matchings constructing*

R.R. Kamalian

Institute for Informatics and Automation Problems of National Academy of Sciences of Armenia, 0014, Armenia

Ijevan Branch of Yerevan State University

Armenian-Russian State University

e-mail: rrkamalian@yahoo.com

and

V. V. Mkrtchyan

Department of Informatics and Applied Mathematics, Yerevan State University, 0025, Armenia

Institute for Informatics and Automation Problems of National Academy of Sciences of Armenia, 0014, Armenia

e-mail: vahanmkrtchyan2002@{ysu.am, ipia.sci.am, yahoo.com}

The author is supported by a grant of Armenian National Science and Educational Fund

* The paper presents the second author's Master thesis defended under supervision of the first author in Yerevan State University in May of 2003

**Abstract**

For bipartite graphs the *NP*-completeness is proved for the problem of existence of maximum matching which removal leads to a graph with given lower(upper) bound for the cardinality of its maximum matching.

**Introduction**

We consider finite undirected graphs $G = (V(G), E(G))$ without multiple edges or loops [17], where $V(G)$ and $E(G)$ are the sets of vertices and edges of $G$, respectively. The maximum degree of a vertex of $G$ is denoted by $\Delta(G)$. The cardinality of a maximum matching [17] of $G$ is denoted by $\beta(G)$.

Let $X = \{x_1, \ldots, x_n\}$ be a set of the boolean variables. A 2-clause $C_j$ is a disjunction of the following form $C_j = l_{j_{i_1}} \vee l_{j_{i_2}}$, in which $j_{i_1} < j_{i_2}$ and $l_{j_{i_p}}$ is a literal of $x_{j_{i_p}}$, i.e., $l_{j_{i_p}}$ is either $x_{j_{i_p}}$ or $\bar{x}_{j_{i_p}}$. In [12, 13] the *NP*-completeness of the following problem is shown:

**PROBLEM 1.** Max *E*2- SAT.

**CONDITION.** Given a set $X = \{x_1, \ldots, x_n\}$ of the boolean variables, a set $C = \{C_1, \ldots, C_m\}$ of $m$ distinct 2-clauses, and a positive integer $K$, $K \leq m$.

**QUESTION.** Does there exist an assignment of variables $x_1,\ldots,x_n$ which satisfies at least $K$ clauses?

Note that without loss of generality, we may always assume that for each instance of Max E2- SAT and for each variable there are at least two clauses containing a literal of the variable.

Consider the following two problems:

**PROBLEM 2.**
**CONDITION.** Given a graph $G$ and a positive integer $k$.
**QUESTION.** Does there exist a maximum matching $F$ of $G$ such that $\beta(G\backslash F) \geq k$ ?

**PROBLEM 3.**
**CONDITION.** Given a graph $G$ and a positive integer $k$.
**QUESTION.** Does there exist a maximum matching $F$ of $G$ such that $\beta(G\backslash F) \leq k$ ?

A well-investigated problem related to matchings is the problem of construction of a maximum matching in a graph. The criterion for maximality of a matching is found in [3, 30]. Historically, the first algorithm constructing a maximum matching for arbitrary graphs was presented in [7] and had a $O(|V|^4)$ runtime. Thereafter, more effective algorithms were found which ran in time $O(|V|^{\frac{1}{2}} \cdot |E|)$ [28] and $O(|V|^{2.5})$ [10]. Many polynomial algorithms are presented for solving mentioned problem in various classes of graphs [1, 6,14, 19, 26].

The problems, which require to construct not just a maximum matching but one which satisfies some additional properties, are also of remarkable importance.

The classical "Assignment problem" is such a one [8, 9, 22-24]. As another example, it can be noted that in a stage of Christofides's algorithm for solving "Traveling Salesman Problem" with the inequality of a triangle, construction of a minimum weighted perfect matching [8] in a complete graph on an even number of vertices is needed. For mentioned problems polynomial algorithms are found [2, 8, 9, 22-25].Related with coloured matchings in bipartite graphs interesting investigations are carried out in [5], which have not only valuable theoretical importance, but also applications in the problems of existence and construction of timetables. An important class of problems related with matchings is formulated in [15]. For a graph property $P$, a $P$-matching is a set $M$ of disjoint edges such that the subgraph induced by the vertices incident to $M$ has property $P$. For different values of $P$ interesting results are obtained in [4, 16]. In [15] general properties of $P$-matchings are investigated, and for some particular cases of $P$ (such as being acyclic or disconnected) interesting corollaries are noticed. Other problems related with matchings are also considered in [11, 20, 29, 31, 32].

Despite this, the influence of a maximum matching of the given graph on the graph obtained by its removal is not sufficiently investigated. It is clear that graphs obtained from a graph by removing its two, different maximum matchings, may have different values of parameters such as maximum degree, chromatic index, cardinality of a minimum vertex covering of a graph. This approach leads to the problem of constructing a maximum matching of a graph which maximizes or minimizes the value of a parameter that interests us for the graph obtained from original one by removing that matching. In this paper the cardinality of a maximum matching of a graph is taken as such a parameter.

The objective of this paper is the investigation of complexity of constructing this kind of maximum matchings for the class of bipartite graphs. Note that if we remove the requirement of bipartiteness then the *NP*-completeness of Problem 2 follows from [18] since the chromatic class of a cubic graph $G$ equals three if and only if $G$ contains a perfect matching $F$ with $\beta(G \backslash F) = \frac{|V(G)|}{2}$.

On the other hand, if for a positve integer $n$ we denote the simple path and cycle containing $n$ edges by $P_n$ and $C_n$, respectively, then

$k - 1 \leq \beta(P_{2k} \backslash F) \leq k$ for every maximum matching $F$ of the path $P_{2k}, k \geq 1$, and there are maximum matchings $F'$ and $F''$ of the path $P_{2k}, k \geq 2$, such that $\beta(P_{2k} \backslash F') = k - 1$, $\beta(P_{2k} \backslash F'') = k$;

$\beta(P_{2k+1} \backslash F) = k$ for the only maximum matching $F$ of the path $P_{2k+1}, k \geq 1$,

$\beta(C_n \backslash F) = \left[\frac{n}{2}\right]$ for every maximum matching $F$ of the cycle $C_n$.

This observation immediately implies that the Problems 2 and 3 are polynomially solvable in the class of graphs $G$ satisfying $\Delta(G) \leq 2$.

Moreover, the problems are also polynomially solvable in the class of regular bipartite graphs, since for every perfect matching $F$ of the *r*-regular bipartite graph $G$ ($r \geq 2$) we have $\beta(G \backslash F) = \frac{|V(G)|}{2}$.

Finally, let us also note that in [21] two polynomial algorithms are presented for solving the optimization versions of the Problems 2 and 3 in the class of trees.

Non-defined terms and conceptions can be found at [12, 17, 27, 33].

**Main results**

**Theorem 1.** The **PROBLEM 2** is *NP*-complete for connected bipartite graphs $G$ with $\Delta(G) = 3$.

**Proof.** Evidently [10, 28], the **PROBLEM 2** belongs to *NP*. Let us describe a polynomial algorithm which reduces the **PROBLEM 1** to the **PROBLEM 2** restricted to connected bipartite graphs $G$ with $\Delta(G) = 3$.

Consider an individual problem $I = (X, C, K)$ of the **PROBLEM 1**, in which $X = \{x_1, \ldots, x_n\}$ is the set of the boolean variables, $C = \{C_1, \ldots, C_m\}$ is the set of 2-clauses and $K$ is the positive integer.

Suppose that a rectangular coordinate system is defined on a plane, and consider a variable $x_i, 1 \leq i \leq n$ and a clause $C_j, 1 \leq j \leq m$, containing a literal of $x_i$. Define a graph corresponding to the pair $(x_i, C_j)$ as follows:

if $x_i \in C_j$ then the graph is the following:

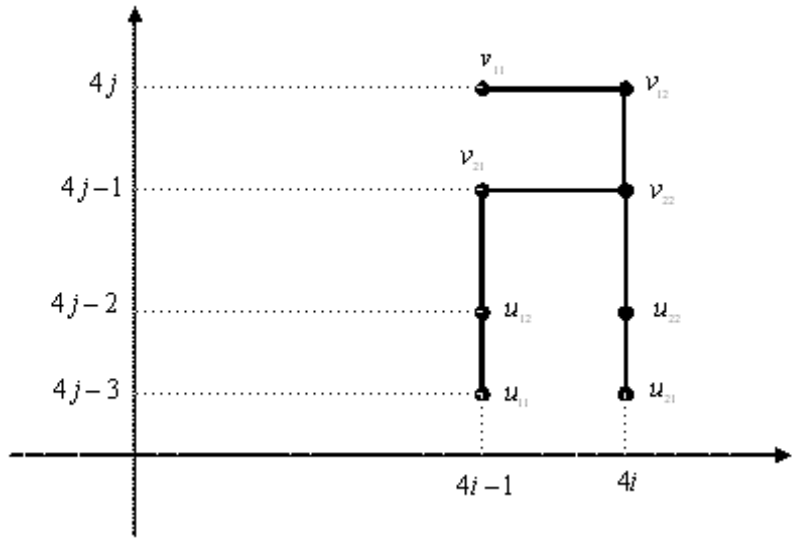

Fig. 1a

if $\bar{x}_i \in C_j$ then the graph is the following:

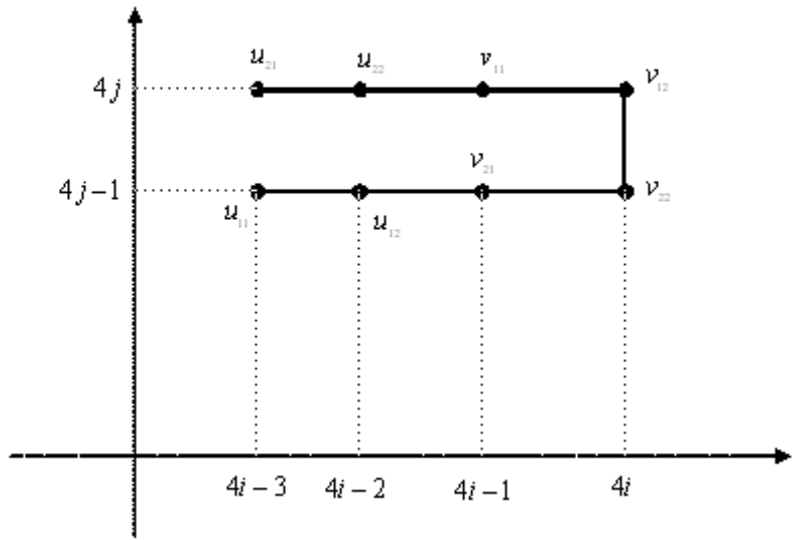

Fig. 1b

Sometimes we will prefer not to draw the whole graph, but to use a conventional sign instead. The conventional sign that will be used for the two graphs shown above is the following:

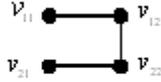

Let us note that the four vertices shown in the figure are the vertices $v_{11}, v_{12}, v_{21}, v_{22}$ of the corresponding graph. Now, if $C_j = l_{j_{i_1}} \vee l_{j_{i_2}}$, ($j_{i_1} < j_{i_2}$) then define the graph $G(C_j)$ corresponding to the clause $C_j$ as follows:

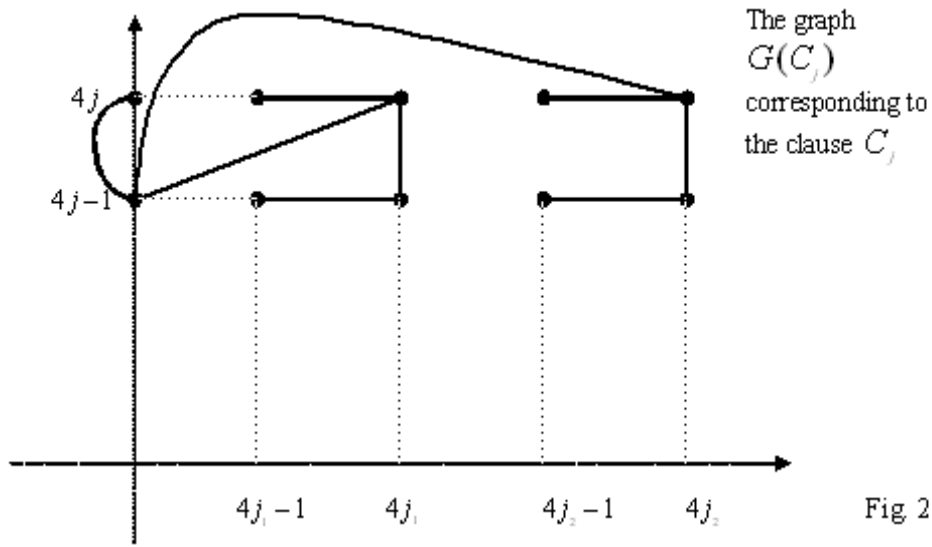

The graph $G(C_j)$ corresponding to the clause $C_j$

Fig. 2

For $i = 1, \ldots, n$ let $C_{j_{i_1}}, C_{j_{i_2}}, \ldots, C_{j_{i_{r(i)}}}$ ($r(i) \geq 2$, $j_{i_1} < j_{i_2} < \ldots < j_{i_{r(i)}}$) denote the clauses containing a literal of $x_i$. Now let us construct the graph $G(I)$ from $G(C_1), G(C_2), \ldots, G(C_m)$ in the following way: for each $i$, $i = 1, \ldots, n$, cyclically connect the subgraphs of $G(C_{j_{i_1}}), G(C_{j_{i_2}}), \ldots, G(C_{j_{i_{r(i)}}})$ that correspond to pairs $(x_i, C_{j_{i_1}}), \ldots, (x_i, C_{j_{i_{r(i)}}})$, respectively, (see fig. 3 below).

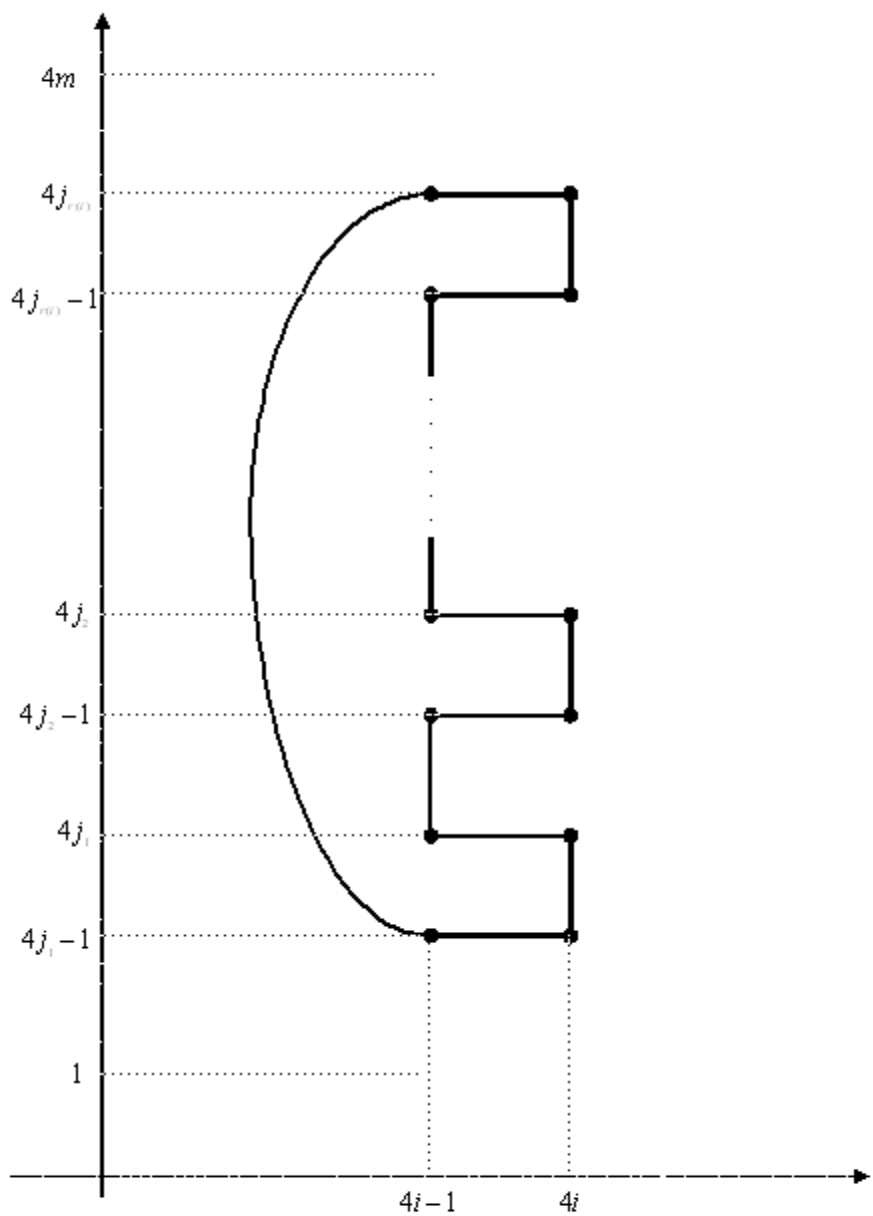

Fig. 3

Note that the graph $G(I)$ may not be connected, therefore, in order to complete the reduction, consider the graph $G_I$ which is constructed from $G(I)$ in the following way (fig. 4):

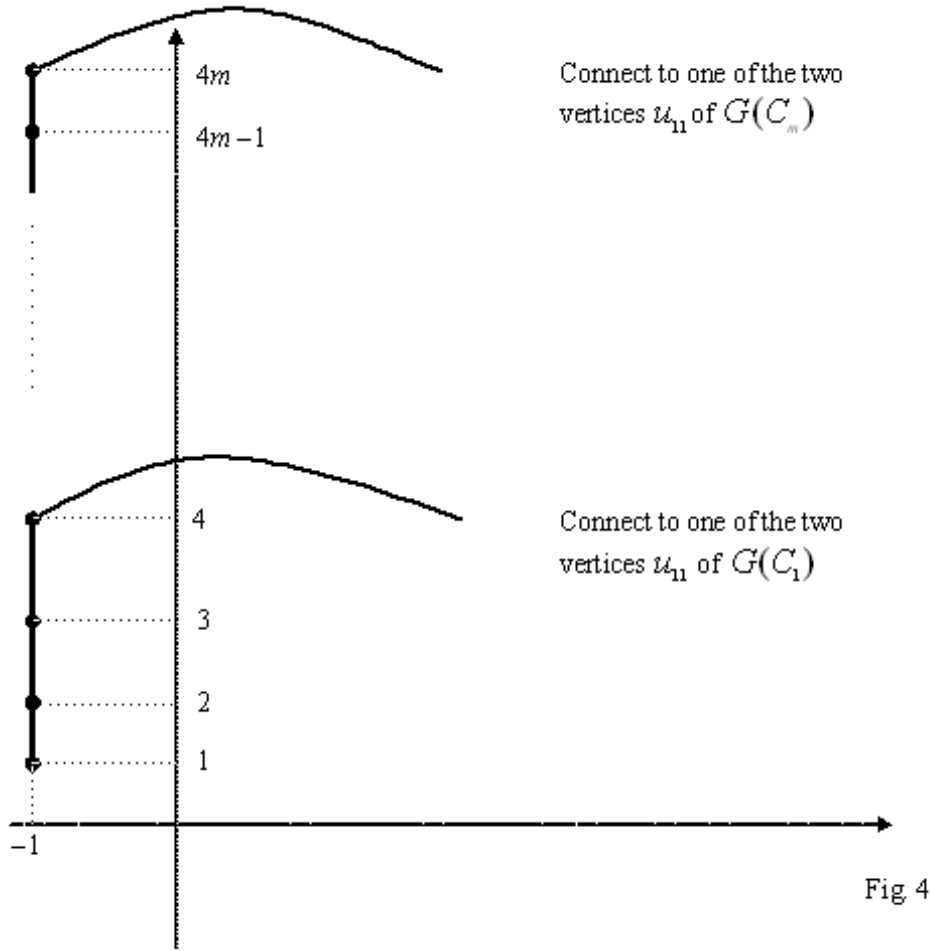

Fig. 4

The description of the graph $G_I$ is completed. Note that $|V(G_I)| = 22m$, $|E(G_I)| = 24m - 1$, $\Delta(G_I) = 3$, $\beta(G_I) = \frac{|V(G_I)|}{2} = 11m$, and $G_I$ is a connected graph. On the other hand, since for every edge $((x,y),(x',y')) \in E(G_I)$ the numbers $|x - x'|$ and $|y - y'|$ have different parity, the sets

$$V_i \equiv \{(x,y) \in V(G_I)/(x+y) \equiv i \pmod 2\}, \ i = 0, 1,$$

form a bipartition of $V(G_I)$, thus $G_I$ is bipartite.

Consider an individual problem $I' = (G_I, k)$ of the **PROBLEM 2**, where $k = 7m + K - 1$. Clearly, $I'$ is constructed from $I$ in a polynomial time.

We claim that $I$ has a positive answer if and only if $I'$ has a positive answer.

Let $I$ have a positive answer, and let $\varepsilon = (\varepsilon_1, \ldots, \varepsilon_n)$ be an assignment which satisfies $K' \geq K$ clauses. For $i = 1, \ldots, n$ let $S_i$ be the subgraph of $G_I$ induced by all vertices $v_{11}, v_{12}, v_{21}, v_{22}$ corresponding to the variable $x_i$ (Fig. 3). Note that $S_i$ is a simple cycle of the length multiple to four.

Consider a perfect matching $F_\varepsilon$ of the graph $G_I$ defined by the following rules:
add edges $((-1, 2j-1), (-1, 2j))$, $j = 1, \ldots, 2m$ to $F_\varepsilon$;
add edges $((0, 4j-1), (0, 4j))$, $j = 1, \ldots, m$ to $F_\varepsilon$;

add all edges $(u_{11}, u_{12}), (u_{21}, u_{22})$ to $F_\varepsilon$;
add all horizontal edges of the cycle $S_i$ to $F_\varepsilon$ if $\varepsilon_i = 0$, $i = 1,\ldots,n$;
add all vertical edges of the cycle $S_i$ to $F_\varepsilon$ if $\varepsilon_i = 1$, $i = 1,\ldots,n$.

Note that for each $j, 1 \leq j \leq m$ the edges of $F_\varepsilon$ which belong to the graph $G(C_j)$ form a matching $F_\varepsilon(j)$ of $G(C_j)$. Moreover,

$$\beta(G(C_j)\backslash F_\varepsilon(j)) = \begin{cases} 6, & \text{if } \varepsilon \text{ satisfies } C_j, \\ 5, & \text{otherwise,} \end{cases}$$

hence,

$$\beta(G_I\backslash F_\varepsilon) = 2m - 1 + 6K' + 5(m - K') \geq 7m - 1 + K = k.$$

Conversely, let $F$ be a perfect matching of the graph $G_I$ satisfying the inequality $\beta(G_I\backslash F) \geq k$. Of course, the edges $((-1, 2j-1), (-1, 2j))$, $j = 1,\ldots, 2m$, $((0, 4j-1), (0, 4j))$, $j = 1,\ldots, m$ and all $(u_{11}, u_{12}), (u_{21}, u_{22})$ belong to $F$. Consider an assignment $\varepsilon_F = (\varepsilon_1(F),\ldots, \varepsilon_n(F))$ where for $i = 1,\ldots, n$

$$\varepsilon_i(F) \equiv \begin{cases} 0, & \text{if } F \text{ takes horizontal edges of } S_i, \\ 1, & \text{otherwise.} \end{cases}$$

Let $K''$ denote the number of clauses satisfied by $\varepsilon_F$. We only need to show that $K'' \geq K$. Note that for $j = 1,\ldots, m$ the edges of $F$ which belong to $E(G(C_j))$ form a matching $F_j$ of $G(C_j)$. Moreover,

$\varepsilon_F$ satisfies $C_j$ if $\beta(G(C_j)\backslash F_j) = 6$;
$\varepsilon_F$ does not satisfy $C_j$ if $\beta(G(C_j)\backslash F_j) = 5$;

hence

$$2m - 1 + 6K'' + 5(m - K'') = \beta(G_I\backslash F) \geq k = 7m - 1 + K,$$

or

$$K'' \geq K.$$

Proof of the **Theorem 1** is complete.

**Theorem 2.** The **PROBLEM 3** is *NP*-complete for connected bipartite graphs $G$ with $\Delta(G) = 3$.

**Proof.** Evidently [10, 28], the **PROBLEM 3** belongs to *NP*. Let us describe a polynomial algorithm which reduces the **PROBLEM 1** to the **PROBLEM 3** restricted to connected bipartite graphs $G$ with $\Delta(G) = 3$.

Consider an individual problem $I = (X, C, K)$ of the **PROBLEM 1**, in which $X = \{x_1,\ldots, x_n\}$ is the set of the boolean variables, $C = \{C_1,\ldots, C_m\}$ is the set of 2-clauses and $K$ is the positive integer.

Again, we will assume that a rectangular coordinate system is defined on a plane. Consider a variable $x_i, 1 \leq i \leq n$ and a clause $C_j, 1 \leq j \leq m$, containing a literal of $x_i$. Define a graph corresponding to the pair $(x_i, C_j)$ in the same way as we did in the proof of the **Theorem 1** (fig. 1a and fig. 1b). Here we will also use the same conventional sign for these graphs.

Now, if $C_j = l_{j_{i_1}} \vee l_{j_{i_2}}$, ($j_{i_1} < j_{i_2}$) then define the graph $G(C_j)$ corresponding to the clause $C_j$ as follows:

if $l_{j_{i_2}} = x_{j_{i_2}}$ then the graph is the following:

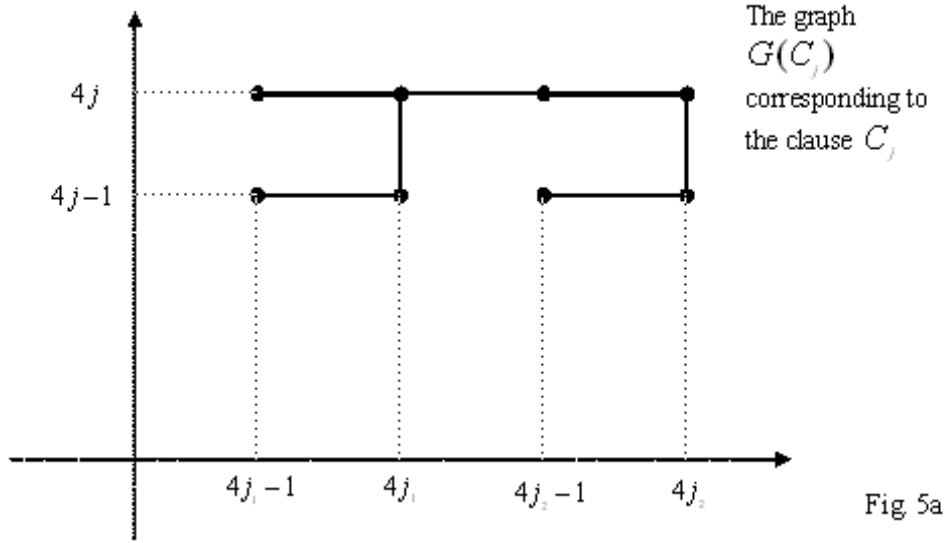

Fig. 5a

if $l_{j_{i_2}} = \bar{x}_{j_{i_2}}$ then the graph is the following:

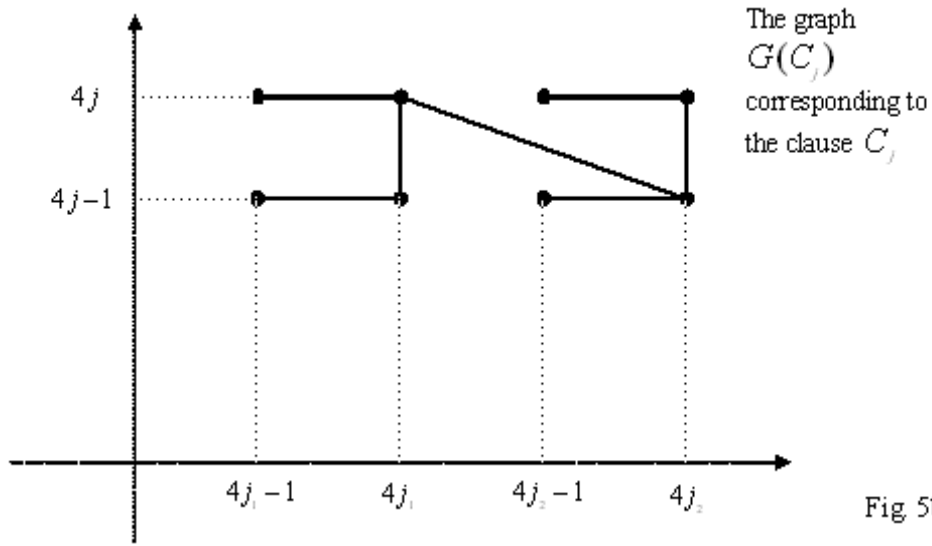

Fig. 5b

Define the graphs $G(I)$ and $G_I$ in the same way as we did in fig. 3 and fig. 4 taking into account that in this case the graphs $G(C_j)$ $j = 1,\ldots,m$ should be understood in sense of fig. 5a and fig 5b.

The description of the graph $G_I$ is completed. Note that $|V(G_I)| = 20m$, $|E(G_I)| = 22m - 1$, $\Delta(G_I) = 3$, $\beta(G_I) = \frac{|V(G_I)|}{2} = 10m$, and $G_I$ is a connected graph. Again by

the construction of $G_I$ the sets
$$V_i = \{(x,y) \in V(G_I)/(x+y) \equiv i \pmod 2\}, \ i = 0, 1,$$
form a bipartition of $V(G_I)$, thus $G_I$ is bipartite.

Consider an individual problem $I' = (G_I, k)$ of the **PROBLEM 3**, where $k = 8m - 1 - K$. Clearly, $I'$ is constructed from $I$ in a polynomial time.

We claim that $I$ has a positive answer if and only if $I'$ has a positive answer.

Let $I$ have a positive answer, and let $\varepsilon = (\varepsilon_1, \ldots, \varepsilon_n)$ be an assignment which satisfies $K' \geq K$ clauses. For $i = 1, \ldots, n$ let $S_i$ be the subgraph of $G_I$ induced by all vertices $v_{11}, v_{12}, v_{21}, v_{22}$ corresponding to the variable $x_i$ (fig. 3). Note that $S_i$ is a simple cycle of the length multiple to four.

Consider a perfect matching $F_\varepsilon$ of the graph $G_I$ defined by the following rules:

add edges $((-1, 2j-1), (-1, 2j))$, $j = 1, \ldots, 2m$ to $F_\varepsilon$;

add all edges $(u_{11}, u_{12}), (u_{21}, u_{22})$ to $F_\varepsilon$;

add all horizontal edges of the cycle $S_i$ to $F_\varepsilon$ if $\varepsilon_i = 1$, $i = 1, \ldots, n$;

add all vertical edges of the cycle $S_i$ to $F_\varepsilon$ if $\varepsilon_i = 0$, $i = 1, \ldots, n$.

Note that for each $j, 1 \leq j \leq m$ the edges of $F_\varepsilon$ which belong to the graph $G(C_j)$ form a matching $F_\varepsilon(j)$ of $G(C_j)$. Moreover,

$$\beta(G(C_j) \backslash F_\varepsilon(j)) = \begin{cases} 5, & \text{if } \varepsilon \text{ satisfies } C_j, \\ 6, & \text{otherwise,} \end{cases}$$

hence,
$$\beta(G_I \backslash F_\varepsilon) = 2m - 1 + 5K' + 6(m - K') \leq 8m - 1 - K = k.$$

Conversely, let $F$ be a perfect matching of the graph $G_I$ satisfying the inequality $\beta(G_I \backslash F) \leq k$. Of course, the edges $((-1, 2j-1), (-1, 2j))$, $j = 1, \ldots, 2m$, and all $(u_{11}, u_{12}), (u_{21}, u_{22})$ belong to $F$. Consider an assignment $\varepsilon_F = (\varepsilon_1(F), \ldots, \varepsilon_n(F))$ where for $i = 1, \ldots, n$

$$\varepsilon_i(F) \equiv \begin{cases} 1, & \text{if } F \text{ takes horizontal edges of } S_i, \\ 0, & \text{otherwise.} \end{cases}$$

Let $K''$ denote the number of clauses satisfied by $\varepsilon_F$. We only need to show that $K'' \geq K$. Note that for $j = 1, \ldots, m$ the edges of $F$ which belong to $E(G(C_j))$ form a matching $F_j$ of $G(C_j)$. Moreover,

$$\varepsilon_F \text{ satisfies } C_j \text{ if } \beta(G(C_j) \backslash F_j) = 5;$$
$$\varepsilon_F \text{ does not satisfy } C_j \text{ if } \beta(G(C_j) \backslash F_j) = 6;$$

hence
$$2m - 1 + 5K'' + 6(m - K'') = \beta(G_I \backslash F) \leq k = 8m - 1 - K,$$

or
$$K'' \geq K.$$

Proof of the **Theorem 2** is complete.

**Acknowledgement**. We thank our referees for careful reading the earlier versions of this paper and for valuable suggestions that helped us to improve the paper significantly.